# Analytical investigation on the minimum traffic delay at a two-phase intersection considering the dynamical evolution process of queues

Hong-Ze Zhang[1], Rui Jiang[1,2], Mao-Bin Hu[1], Bin Jia[2]

[1]School of Engineering Science, University of Science and technology of China, Hefei 230026, China

[2]MOE Key Laboratory for Urban Transportation Complex Systems Theory and Technology, Beijing Jiaotong University, Beijing 100044, China

Abstract: This paper has studied the minimum traffic delay at a two-phase intersection, taking into account the dynamical evolution process of queues. The feature of delay function has been studied, which indicates that the minimum traffic delay must be achieved when equality holds in at least one of the two constraints. We have derived the minimum delay as well as the corresponding traffic signal period, which shows that two situations are classified. Under certain circumstance, extra green time is needed for one phase while otherwise no extra green time should be assigned in both phases. Our work indicates that although the clearing policies were shown in many experiments to be optimal at isolated intersections, it is sometimes not the case.

1. Introduction

Traffic congestion has been a serious problem all over the world, which has attracted the wide attention of scientists [1-6]. In uninterrupted traffic flow on highways and expressways, recurrent traffic congestion is usually induced by bottlenecks such as merges and diverges. In urban traffic, the delays are usually due to stop at signalized intersections [7].

The queue theory has been applied to study traffic delay at the signalized intersection for decades. For example, Webster [8] has derived the analytical expressions for delays and queues under the assumption that vehicle arrivals follow Poisson distribution. Miller [9] and Newell [10] proposed analytical models to study residual queues that are not able to dissipate at the end of green phase.

For time-dependent traffic demands, the queue model has also been used to calculate average delay for fixed-time signal control and adaptive traffic signal control strategy. For example, Newell [11] proposed a clearing strategy based on rolling horizon scheme, in which a signal serves a two-phase intersection with one-way traffic streams alternatively. In this strategy, the cyclic process repeats as follows: The first stream is served until the queue dissipates, then the signal serves the second stream until the queue dissipates, then the signal switches back to serve the first stream. Mirchandani and Zou [12] have developed a numerical algorithm to compute steady-state performance measures such as average delays and expected queue lengths in Newell's strategy, which are in agreement with simulation based results. Recently a max pressure control method has been developed for a network of signalized intersections [13], which is claimed to be able to maximize network throughput.

However, in these studies, dynamical evolution process of the queue has not been considered. The queue is a point queue, in which the basic assumption is that when the arrival rate is larger (smaller) than the service rate, the queue length increases (decreases). Actually, when traffic signal turns into green, the queue dissipates and at the same time, its downstream front propagates upstream.

To study traffic flows at the intersections considering the effect of physical queues, many works have been carried out based on simulation frameworks, see e.g., [14-18]. However, analytical investigations of traffic delays are absent. Recently Helbing [19] has derived the average delay of a traffic stream at a traffic signal, taking into account the dynamical evolution process of the queue. Based on Helbing's work, this paper aims to derive the minimum traffic delay at a two-phase intersection with two traffic streams.

The paper is organized as follows. In section 2, we review the analytical results of traffic delay of a traffic stream at a traffic signal. Section 3 derives the minimum traffic delay and the corresponding traffic period at a two-phase intersection with two one-way traffic streams. Section 4 generalizes the investigation to a two-phase intersection with two two-way traffic streams. Conclusion is given in section 5.

2. Average delay of a traffic stream

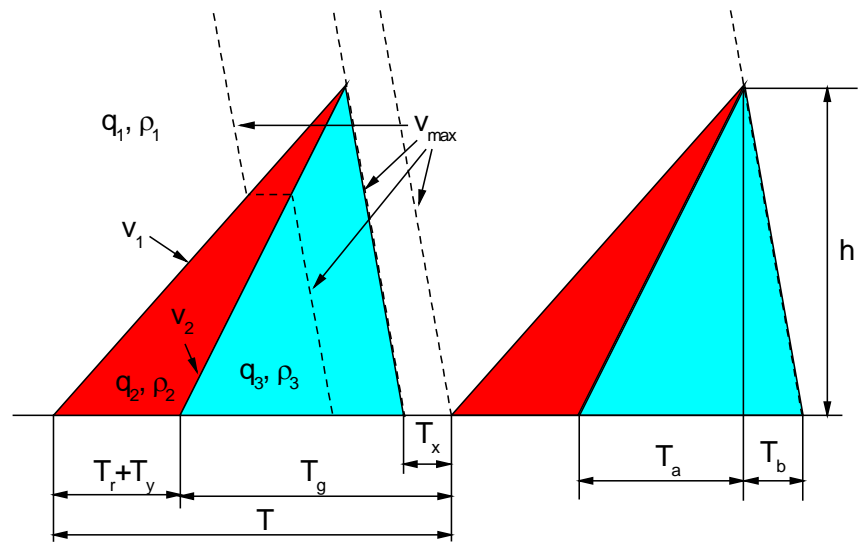

Fig.1 Sketch of the dynamical evolution process of the queue (denoted as red). The cyan denotes saturated outflow from the queue. The dashed lines show typical trajectories of some vehicles.

Fig.1 shows dynamical evolution process of the queue at traffic signal. The arrival flow rate is assumed to be constant. When traffic light is in red, the queue length increases. The propagating speed of the shock front separating the queue and the arrival flow can be calculated by the R-H condition

$$v_1 = \left|\frac{q_1 - q_2}{\rho_1 - \rho_2}\right| \tag{1}$$

where q and ρ are flow rate and density, respectively. The subscript 1 and 2 denote "upstream" and "downstream" of the shock, respectively. Obviously, downstream of the shock, the flow rate in the queue $q_2 = 0$ and the density $\rho_2 = \rho_{jam}$. Here $\rho_{jam}$ is jam density. Furthermore, we assume that the arrival flow $q_1 = \rho_1 v_{max}$, with $v_{max}$ the maximum velocity of traffic flow. Therefore, the shock velocity

$$v_1 = \left|\frac{q_1}{q_1/v_{max} - \rho_{jam}}\right| = \frac{q_1}{\rho_{jam} - q_1/v_{max}} \tag{2}$$

When the traffic signal turns into green, the queue begins to dissipate. The propagating speed of the dissipation front can also be calculated by R-H condition

$$v_2 = \left|\frac{q_3 - q_2}{\rho_3 - \rho_2}\right| \tag{3}$$

Here the subscript 2 and 3 denote upstream and downstream of the dissipation front, respectively. We assume that the outflow from the queue is saturated flow so that $q_3 = q_{max} = \rho_{max} v_{max}$ and $\rho_3 = \rho_{max}$. Here $q_{max}$ is saturated flow rate. Therefore,

$$v_2 = \frac{q_{max}}{\rho_{jam} - q_{max}/v_{max}} \tag{4}$$

When the traffic signal turns into yellow, vehicles that have not crossed the stop line need to top. Therefore, it can be regarded as an equivalent red signal.

From Fig.1, one has

$$h = v_{max} T_b$$
$$h = v_2 T_a$$
$$h = v_1 (T_r + T_y + T_a)$$

Therefore, $T_a$, $T_b$ and $h$ can be solved

$$h = \frac{T_r + T_y}{\frac{1}{v_1} - \frac{1}{v_2}}$$

$$T_a = \frac{1}{v_2}\left(\frac{T_r + T_y}{\frac{1}{v_1} - \frac{1}{v_2}}\right)$$

$$T_b = \frac{1}{v_{max}}\left(\frac{T_r + T_y}{\frac{1}{v_1} - \frac{1}{v_2}}\right)$$

where $T_g$ is green signal time, $T_r$ is red signal time, $T_y$ is yellow signal time, $T = T_g + T_r + T_y$ is one traffic period.

The average delay of the delayed vehicles can be easily calculated as $\frac{T_r + T_y}{2}$. During one traffic period, the number of delayed vehicles is $q_1(T_r + T_y + T_a + T_b)$, the total number of arrival

vehicles is $q_1 T$. Therefore, the average delay

$$de = \frac{q_1(T_r+T_y+T_a+T_b)\frac{T_r+T_y}{2}}{q_1(T_g+T_r+T_y)} \quad (6)$$

3. Two-phase intersection with two one-way streams

In this section we derive the minimum traffic delay at a two-phase intersection with two one-way traffic streams. The demands of the two streams are $q_1$ and $q_1^{'}$. We fix the time $T_y$. The average delay of the vehicles of the two streams is

$$de = \frac{q_1(T_r+T_y+T_a+T_b)\frac{T_r+T_y}{2}+q_1^{'}(T_r^{'}+T_y+T_a^{'}+T_b^{'})\frac{T_r^{'}+T_y}{2}}{(q_1+q_1^{'})(T_g+T_r+T_y)} \quad (7)$$

here the superscript ' denotes variables corresponding to traffic stream $q_1^{'}$. To derive the minimum value of traffic delay, the two constraints

$$T_g \geq T_a + T_b \quad (8)$$
$$T_g^{'} \geq T_a^{'} + T_b^{'} \quad (9)$$

need to be considered. Otherwise traffic congestion will emerge.

The average delay in Eq.(7) can be reformulated as

$$de = de(T, T_g) = \frac{m(T-T_g)^2 + m^{'}(T_g+2T_y)^2}{T} \quad (10)$$

Here

$$m = \frac{q_1(1+p)}{2(q_1+q_1^{'})}$$

$$m^{'} = \frac{q_1^{'}(1+p^{'})}{2(q_1+q_1^{'})}$$

$$p = \left(\frac{1}{v_2}+\frac{1}{v_{max}}\right)\left(\frac{1}{\frac{1}{v_1}-\frac{1}{v_2}}\right)$$

$$p^{'} = \left(\frac{1}{v_2}+\frac{1}{v_{max}}\right)\left(\frac{1}{\frac{1}{v_1^{'}}-\frac{1}{v_2}}\right)$$

In Eq.(10), there are only two variables T and $T_g$.

*3.1. Feature of delay function*

To derive the minimum value of delay, we firstly study the feature of function de(T,$T_g$). Taking second derivative to $T_g$, one has

$$\frac{\partial de^2}{\partial^2 T_g} = \frac{2(m+m^{'})}{T} \quad (11)$$

which is obviously larger than zero. Now we take first derivative of de(T,$T_g$) to $T_g$, which yields

$$\frac{\partial de}{\partial T_g} = \frac{2m(T_g-T)+2m^{'}(T_g+2T_y)}{T}$$

Let $\frac{\partial de}{\partial T_g} = 0$, we can obtain

$$T_g = \frac{mT - 2m'T_y}{m+m'} \quad (12)$$

Substituting Eq.(12) into Eq.(10), one has

$$de = de(T) = \frac{mm'(T+2T_y)^2}{(m+m')T} \quad (13)$$

which is the minimum delay at given T. We take first derivative of de (T) to T, which yields

$$\frac{d(de)}{dT} = \frac{mm'(T^2 - 4T_y^2)}{(m+m')T^2} \quad (14)$$

Obviously it is larger than zero, since $T = T_g + T_g' + 2T_y > 2T_y$. This means that the minimum value of traffic delay decreases with the decrease of T. Based on the feature of delay function $de(T, T_g)$, one knows that the minimum traffic delay must be achieved when equality holds in at least one of the two constraints (8) and (9).

*3.2. Minimum delay*

3.2.1 *Case 1*: $T_g = T_a + T_b$

Without loss of generality, we suppose $q_1 \leq q_1'$. Let us firstly study the case $T_g = T_a + T_b$. In this case, one can easily derive that

$$T = \frac{p+1}{p} T_g \quad (15)$$

Substituting Eq.(15) into Eq.(10), one has

$$de(T_g) = \alpha T_g + \beta \frac{(T_g + 2T_y)^2}{T_g} \quad (16)$$

Here

$$\alpha = \frac{q_1}{2(q_1 + q_1')p}$$

$$\beta = \frac{pq_1'(1+p')}{2(q_1 + q_1')(1+p)}$$

Substituting Eq.(15) into (9), the constraint becomes

$$T_g \geq \frac{2p(p'+1)T_y}{(1-pp')} \quad (17)$$

From the two constraints (8) and (9), one can easily derive, respectively

$$\left(1 + \frac{1}{p}\right) T_g \geq T$$

$$(1+p')(T_g + 2T_y) \leq T$$

which leads to

$$(1+p')(T_g + 2T_y) \leq \left(1 + \frac{1}{p}\right) T_g$$

A reformulation yields

$$(1-pp') \geq \frac{2p(1+p')T_y}{T_g}$$

Therefore, $1 - pp' > 0$.

Now we take first derivative of $de(T_g)$ to $T_g$, which yields

$$\frac{d(de)}{dT_g} = \alpha + \beta - 4\beta \frac{T_y^2}{T_g^2} \tag{18}$$

Let $\frac{d(de)}{dT_g} = 0$, we can solve

$$T_g = 2\sqrt{\frac{\beta}{\alpha+\beta}} T_y \tag{19}$$

Substituting Eq.(19) into constraint (17), one has

$$\sqrt{\frac{\beta}{\alpha+\beta}} \geq \frac{p(p'+1)}{(1-pp')} \tag{20}$$

If Inequality (20) is satisfied, then Eq.(19) is the solution corresponding to the minimum delay. Otherwise, $T_g = \frac{2p(p'+1)T_y}{(1-pp')}$ is the solution corresponding to the minimum delay. Substituting $T_g = \frac{2p(p'+1)T_y}{(1-pp')}$ or $T_g = 2\sqrt{\frac{\beta}{\alpha+\beta}} T_y$ into Eq.(16), the minimum delay

$$(de)_{min} = \frac{(q_1(1+p')+q_1'(1+p))T_y}{(q_1+q_1')(1-pp')} \tag{21}$$

or

$$(de)_{min} = 4(\beta + \sqrt{\beta(\alpha+\beta)})T_y \tag{22}$$

Fig.2 shows three typical plots of $\left(2\sqrt{\frac{\beta}{\alpha+\beta}}T_y - \frac{2p(p'+1)T_y}{(1-pp')}\right)$ at three different values of $q_1$. One can see that when $q_1 > q_{1,c}$, Inequality (20) will never be satisfied. In this case, the solution is always $T_g = \frac{2p(p'+1)T_y}{(1-pp')}$, which corresponds to the situation that $T_x = T_x' = 0$. Namely, the traffic signal needs to be switched to let the last vehicle that has been involved in the queue to cross the stop line. The vehicles that have not involved in the queue need to stop to wait for next green signal.

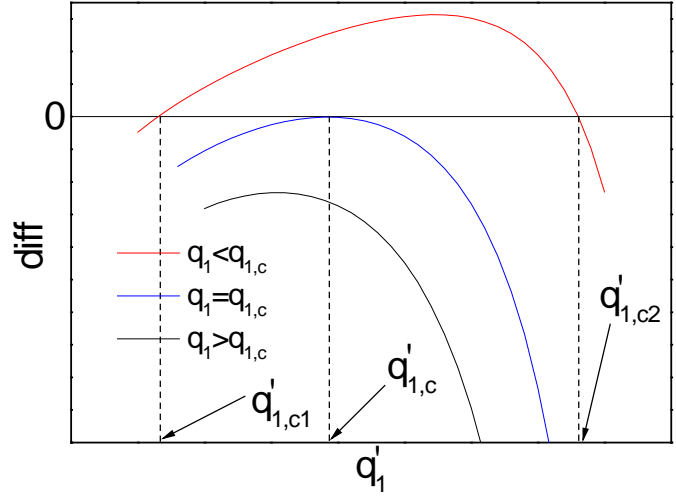

Fig.2 Three typical plots of $\text{diff} = \left(2\sqrt{\frac{\beta}{\alpha+\beta}}T_y - \frac{2p(p'+1)T_y}{(1-pp')}\right)$.

However, when $q_1 < q_{1,c}$, Inequality (20) will be satisfied when $q'_{1,c1} \leq q'_1 \leq q'_{1,c2}$. In this case, the solution is $T_g = 2\sqrt{\frac{\beta}{\alpha+\beta}}T_y$, corresponding to the situation that $T_x = 0$ and

$$T'_x = 2\left(\left(\frac{1}{p} - p'\right)\sqrt{\frac{\beta}{\alpha+\beta}} - (1+p')\right)T_y \geq 0$$

Under this circumstance, for stream $q'_1$, when the last vehicle that has been involved in the queue has crossed the stop line, some extra green time is still needed in order to achieve minimum delay. With the increase of $q_1$, the range $q'_{1,c1} \leq q'_1 \leq q'_{1,c2}$ shrinks. When $q_1 = q_{1,c}$, $q'_{1,c1} = q'_{1,c2} = q'_{1,c}$, see Fig.2.

3.2.2 *Case 2*: $T'_g = T'_a + T'_b$

Next study the case $T'_g = T'_a + T'_b$. In this case, one can easily derive that
$$T = (p'+1)T_g + 2(p'+1)T_y \tag{23}$$
Substituting Eq.(23) into Eq.(10), one has
$$de(T_g) = \tilde{\alpha}\left(p'^2(T_g + 2T_y) + 4p'T_y + \frac{4T_y^2}{T_g+2T_y}\right) + \tilde{\beta}(T_g + 2T_y) \tag{24}$$

Here
$$\tilde{\alpha} = \frac{q_1(p+1)}{2(q_1 + q'_1)(p'+1)}$$
$$\tilde{\beta} = \frac{q'_1}{2(q_1 + q'_1)}$$

Substituting Eq.(23) into (8), the constraint becomes

$$T_g \geq \frac{2p(p'+1)T_y}{(1-pp')} \tag{25}$$

which is the same as constraint (17). Now we take first derivative of $de(T_g)$ to $T_g$, which yields

$$\frac{d(de)}{dT_g} = \tilde{\alpha}p'^2 - \frac{4\tilde{\alpha}T_y^2}{(T_g+2T_y)^2} + \tilde{\beta} \tag{26}$$

Let $\frac{d(de)}{dT_g} = 0$, we can solve

$$T_g = 2T_y\left(\sqrt{\frac{\tilde{\alpha}}{\tilde{\alpha}p'^2+\tilde{\beta}}} - 1\right) \tag{27}$$

Substituting Eq.(27) into constraint (25), one has

$$\sqrt{\frac{\tilde{\alpha}}{\tilde{\alpha}p'^2+\tilde{\beta}}} - 1 \geq \frac{p(p'+1)}{(1-pp')} \tag{28}$$

If Inequality (28) is satisfied, then Eq.(27) is the solution corresponding to the minimum delay. Otherwise, $T_g = \frac{2p(p'+1)T_y}{(1-pp')}$ is the solution corresponding to the minimum delay. Since $q_1 \leq q_1'$, one has $v_1 \leq v_1'$, $p \leq p'$. It can be easily proved that $\sqrt{\frac{\tilde{\alpha}}{\tilde{\alpha}p'^2+\tilde{\beta}}} - 1 < 0$. Thus, Inequality (28) is never met. Therefore, in Case 2, the solution is always $T_g = \frac{2p(p'+1)T_y}{(1-pp')}$.

4. Two-phase intersection with two two-way streams

Now we generalize the investigation to a two-phase intersection with two two-way streams. The demands of the East-to-West and West-to-East streams are denoted as $q_1$ and $q_0$, respectively. The demands of the North-to-South and South-to-North streams are denoted as $q_1'$ and $q_0'$, respectively. Without loss of generality, we assume $q_1 > q_0$, $q_1' > q_0'$, and $q_1' > q_1$. The average delay of the vehicles is thus

$$de = \frac{m(T-T_g)^2 + m'(T_g+2T_y)^2}{T} \tag{29}$$

here

$$m = \frac{q_1(1+p_1) + q_0(1+p_0)}{2(q_1+q_0+q_1'+q_0')}$$

$$m' = \frac{q_1'(1+p_1') + q_0'(1+p_0')}{2(q_1+q_0+q_1'+q_0')}$$

$$p_1 = \left(\frac{1}{v_2} + \frac{1}{v_{max}}\right)\left(\frac{1}{\frac{1}{v_1}-\frac{1}{v_2}}\right)$$

$$p_0 = \left(\frac{1}{v_2} + \frac{1}{v_{max}}\right)\left(\frac{1}{\frac{1}{v_0}-\frac{1}{v_2}}\right)$$

$$p_1' = \left(\frac{1}{v_2} + \frac{1}{v_{max}}\right)\left(\frac{1}{\frac{1}{v_1'}-\frac{1}{v_2}}\right)$$

$$p_0' = \left(\frac{1}{v_2} + \frac{1}{v_{max}}\right)\left(\frac{1}{\frac{1}{v_0'} - \frac{1}{v_2}}\right)$$

$$v_0 = \frac{q_0}{\rho_{jam} - q_0/v_{max}}$$

$$v_0' = \frac{q_0'}{\rho_{jam} - q_0'/v_{max}}$$

Since $q_1 > q_0$, $q_1' > q_0'$, the two constraints are

$$T_g \geq T_{a1} + T_{b1} \tag{30}$$

$$T_g' \geq T_{a1}' + T_{b1}' \tag{31}$$

One can easily prove that the delay function (29) has the same feature as delay function in Eq.(10). Therefore, the minimum delay is achieved when equality holds in at least one of the two constraints (30) and (31).

In the case $T_g = T_{a1} + T_{b1}$, one can derive similar as in section 3.2.1 that $de(T_g) = \alpha T_g + \beta \frac{(T_g + 2T_y)^2}{T_g}$.

If $\sqrt{\frac{\beta}{\alpha+\beta}} \geq \frac{p_1(1+p_1')}{(1-p_1 p_1')}$ is satisfied, then $T_g = 2\sqrt{\frac{\beta}{\alpha+\beta}} T_y$ is the solution corresponding to the minimum delay. Otherwise, $T_g = \frac{2p_1(1+p_1')T_y}{(1-p_1 p_1')}$ is the solution corresponding to the minimum delay. Note that here

$$\alpha = \frac{q_0(1+p_0) + q_1(1+p_1)}{2p_1(1+p_1)(q_1 + q_0 + q_1' + q_0')}$$

$$\beta = \frac{p_1 q_1'(1+p_1') + p_1 q_0'(1+p_0')}{2(1+p_1)(q_1 + q_0 + q_1' + q_0')}$$

In the case $T_g' = T_{a1}' + T_{b1}'$, one can derive that $de(T_g) = \tilde{\alpha}\left(p_1'^2(T_g + 2T_y) + 4p_1'T_y + \frac{4T_y^2}{T_g + 2T_y}\right) + \tilde{\beta}(T_g + 2T_y)$. If $\sqrt{\frac{\tilde{\alpha}}{\tilde{\alpha}p_1'^2 + \tilde{\beta}}} - 1 \geq \frac{p_1(1+p_1')}{(1-p_1 p_1')}$ is satisfied, then $T_g = 2\left(\sqrt{\frac{\tilde{\alpha}}{\tilde{\alpha}p_1'^2 + \tilde{\beta}}} - 1\right)T_y$ is the solution corresponding to the minimum delay. Otherwise, $T_g = \frac{2p_1(1+p_1')T_y}{(1-p_1 p_1')}$ is the solution corresponding to the minimum delay. Note that here

$$\tilde{\alpha} = \frac{q_1(1+p_1) + q_0(1+p_0)}{2(q_1 + q_0 + q_1' + q_0')(p_1' + 1)}$$

$$\tilde{\beta} = \frac{q_1'(p_1' + 1) + q_0'(1+p_0')}{2(q_1 + q_0 + q_1' + q_0')(p_1' + 1)}$$

Also note that different from in section 3.2.2, $\sqrt{\frac{\tilde{\alpha}}{\tilde{\alpha}p_1'^2 + \tilde{\beta}}} - 1 \geq \frac{p_1(1+p_1')}{(1-p_1 p_1')}$ can sometimes be satisfied. Under such circumstance, one has $T_x \geq 0$ and $T_x' = 0$.

5. Conclusion

In this paper, we have studied the minimum traffic delay at a two-phase intersection, taking into account the dynamical evolution process of queues. The feature of delay function has been studied, which indicates that the minimum traffic delay must be achieved when equality holds in at least one of the two constraints. We have derived the minimum delay as well as the corresponding traffic signal period, which shows that two situations are classified. Under certain circumstance, extra green time is needed for one phase while otherwise no extra green time should be assigned in both phases. In the special case of two one-way streams,

(i) When $q_1 < q_{1,c}$ and $q'_{1,c1} \leq q'_1 \leq q'_{1,c2}$, some extra green time is needed for the large stream $q'_1$. For the small stream $q_1$, the extra green time should be zero.

(ii) Otherwise, for both streams, extra green time should be zero.

Our work indicates that although the clearing policies, which continue serving a street until its queue has been fully cleared as proposed in Newell [11], were shown in many experiments to be optimal at isolated intersections [20,21], it is not always the case.

In our future work, the investigation will be generalized to multiphase intersection. Another important task is to consider the situation that traffic demand is changing over time.

Acknowledgements: This work was supported by the National Basic Research Program of China under Grant No.2012CB725404, the Natural Science Foundation of China under Grant Nos. 11422221, 71371175, 71222101.